\documentclass[a4paper,10pt]{article}
\pdfoutput=1
\usepackage{graphicx}
\usepackage{fancyhdr}
\usepackage{changebar}
\usepackage{natbib}
\usepackage{amssymb,amsmath}
\usepackage{url}
\usepackage{color}
\usepackage{longtable}
\usepackage{subfig}
\usepackage{lscape}
\usepackage{rotating}
\usepackage{perpage}
\MakePerPage{footnote}
\usepackage[rmargin=4cm,lmargin=3cm,bmargin=4.5cm,tmargin=3.5cm]{geometry}
\usepackage{setspace}
\sffamily
\usepackage[T1]{fontenc}
\usepackage[utf8x]{inputenc}

\newcommand{\kms}{{\,km\,s$^{-1}$}}
\newcommand{\dgr}{{^\circ}}

\newcommand{\la}{\mathrel{\hbox{\rlap{\hbox{\lower4pt\hbox{$\sim$}}}\hbox{$<$}}}}
\newcommand{\ga}{\mathrel{\hbox{\rlap{\hbox{\lower4pt\hbox{$\sim$}}}\hbox{$>$}}}}

\newcommand{\arcmin}{\hbox{$^\prime$}}
\newcommand{\arcsec}{\hbox{$^{\prime\prime}$}}

\newcommand{\fm}{\hbox{$.\!\!^{\rm m}$}}

\newcommand{\fdg}{\hbox{$.\!\!^\circ$}}
\newcommand{\farcm}{\hbox{$.\mkern-4mu^\prime$}}

\newcommand{\msun}{\,{\rm M}_\odot}

\title{{\bf The Mass Distribution of the Great Attractor\\ as Revealed by a Deep NIR Survey}}  
\author{R.C. Kraan-Korteweg$^{\rm 1}$, I.F. Riad$^{\rm 1}$, P.A. Woudt$^{\rm 1}$, T. Nagayama$^{\rm2}$, K. Wakamatsu$^{\rm 3}$\smallskip \\
\begin{small} $^1$Astronomy Department, ACGC, University of Cape Town, South Africa\end{small}\\
\begin{small} $^2$Department of Astrophysics, Nagoya University, Japan
\end{small}\\
\begin{small} $^3$Faculty of Engineering, Gifu University, Japan
\end{small}
}
\date{~}

\begin{document}

\maketitle

\begin{abstract}

\noindent 
This paper presents the analysis of a deep near-infrared $JHK_s$ imaging
survey ($37.5\Box\dgr$) aimed at tracing the galaxy distribution of
the Great Attractor (GA) in the Zone of Avoidance along the so-called Norma
Wall. The resulting galaxy catalog is complete to {\sl
  extinction-corrected} magnitudes $K_s^o < 14\fm8$ for
extinctions less than $A_K \le 1\fm0$ and star densities below $\log
N_{(K<14.0)} \le 4.72$.  Of the 4360 catalogued galaxies, 99.2\% lie
in the hereby constrained 89.5\% of the survey area. Although the analyzed
galaxy distribution reveals no new major galaxy clusters at the GA distance (albeit some more distant ones), the overall number
counts and luminosity density indicate a clear and surprisingly smooth
overdensity at the GA distance that extends over the whole surveyed
region. A mass estimate of the Norma Wall overdensity derived from (a)
galaxy number counts and (b) photometric redshift distribution gives a
lower value compared to the original prediction by Lynden-Bell et
al. 1988 ($\sim 14$\%), but is consistent with more recent independent
assessments.

\end{abstract}

\section{Introduction}
\noindent
The motivation for this deep near-infrared (NIR) imaging survey along
the Norma Wall (henceforth the Norma Wall Survey, NWS) is described by
Wakamatsu 2011 (these proceedings). It includes a description of the
survey area and the derived NIR ($JHK_s$) positional and photometric
parameters of the 4360 identified galaxies. The main goal of the
project is the mapping of the galaxy distribution in the Great
Attractor (GA) region where extinction and star crowding by the Milky
Way is so severe that neither optical nor NIR whole-sky surveys (like
2MASX; Jarrett et al. 2000; Jarrett 2004) allow an estimate of the GA
mass-density, wheras the deepest systematic HI ZOA surveys to date
(Parkes MB; e.g. Kraan-Korteweg et al. \,2005), which allow full
penetration of the ZOA, remain very shallow ($\la
1$gal$/\Box\dgr$). See Kraan-Korteweg \& Lahav (2000); Kraan-Korteweg
(2005) for a review of the various dedicated multi-wavelength ZOA
surveys.

The earlier optical, NIR and HI ZOA results suggest the GA to consist
of a Great Wall-like structure, that extends over $100\dgr$ on the sky
(from Pavo to Abell S0639); it includes the massive clusters Norma
A3627 (the likely core of this structure), CIZA J1324.7-5736 (CIZA),
Cen-Crux and PKS 1343-601 (see e.g. Radburn-Smith et al. 2006 for
further details).

We used the InfraRed Survey Facility (IRSF) at the 1.4\,m Japanese
telescope in Sutherland (SAAO) to obtain $\sim 2800$ $JHKs$-band
survey images of $7\farcm8 \times 7\farcm8$ to probe the most central,
low-latitude ($|b| < 7\dgr$) region of the Norma Wall. It starts above
the Norma cluster (covered by Skelton et al. 2009) and includes the
CIZA and Cen-Crux clusters on the other side of the ZOA. The exposure
times of 10 min are an optimization between survey depth versus sky
background and star density. The longer integrations times and
particularly the higher resolution of the IRSF ($45\arcsec/$pix)
compared to 2MASX allows the detection of galaxies deeper into the
Milky Way and improved photometry (see also Skelton et al. 2009;
Williams et al. these proceedings). Only 5.5\% of the identified
galaxies have counterparts in 2MASX. Further details are given in Riad
(2010, PhD thesis) and the forthcoming catalogue paper (Nagayama, Riad
et al., in prep).

\section{Completeness as a function of extinction}

The first step in the interpretation of the uncovered large-scale
structures within this survey is a good understanding of the
completeness as a function of foreground extinction and star
density. This was assessed from the cumulative number counts for the
observed $JHK_s$ and foremost for the {\sl extinction-corrected} $J^o,
H^o$ and $K_s^o$. The latter were determined by applying the
derivations of Riad et al. (2010) for absorption effects on their
observed isophotal $JHK_s$ magnitudes and radii.

The analysis finds the NWS galaxy catalog complete for {\sl
  isophotal} magnitudes $J^o=15\fm6$, $H^o=15\fm3$ and
$K_s^o=14\fm8$ where dust obscuration does not exceed
$A_{J,H,K_s}<1$ mag respectively (the apparent limits are
$JHK_s$=$16\fm6, 15\fm8$ and $15\fm4$ respectively). For higher
extinction values this drops significantly. But note that the great
majority (99.2$\%$) of the galaxies in the NWS are found in regions
with dust obscuration below $A_{K_s}<1$ mag. In fact, almost 90$\%$ of
the galaxies have foreground extinction less than $A_{K}=0\fm42$
($A_J=0\fm98$, $A_H=0\fm65$).

The completeness limit also depends on stellar density. Our analysis
indicates that the success rate in identifying and parameterizing
galaxies decreases once star densities reach values above $\log
N_{(K<14.0)} > 4.72$ (due to crowding as well as the increased sky
background). Interestingly, this stellar density contour corresponds
closely to the extinction contour of $A_K < 1\fm0$ but lies a bit
lower below the Galactic Plane ($b \sim -2\dgr$, rather than ($b \sim
-1\fdg5$).

We therefore conclude that the survey is complete for $K^o_s < 14\fm8$
in the regions delimited by $A_K < 1\fm0$ and $N_{(K<14.0)} >
4.72$. This comprises $89\%$ ($\sim 33.2\Box\dgr$) of the survey area
and 99.2\% of the 4360 galaxies.

\section{Assessment of the 2-D distribution}

\begin{figure}
\begin{center}
\includegraphics[width=0.69\textwidth]{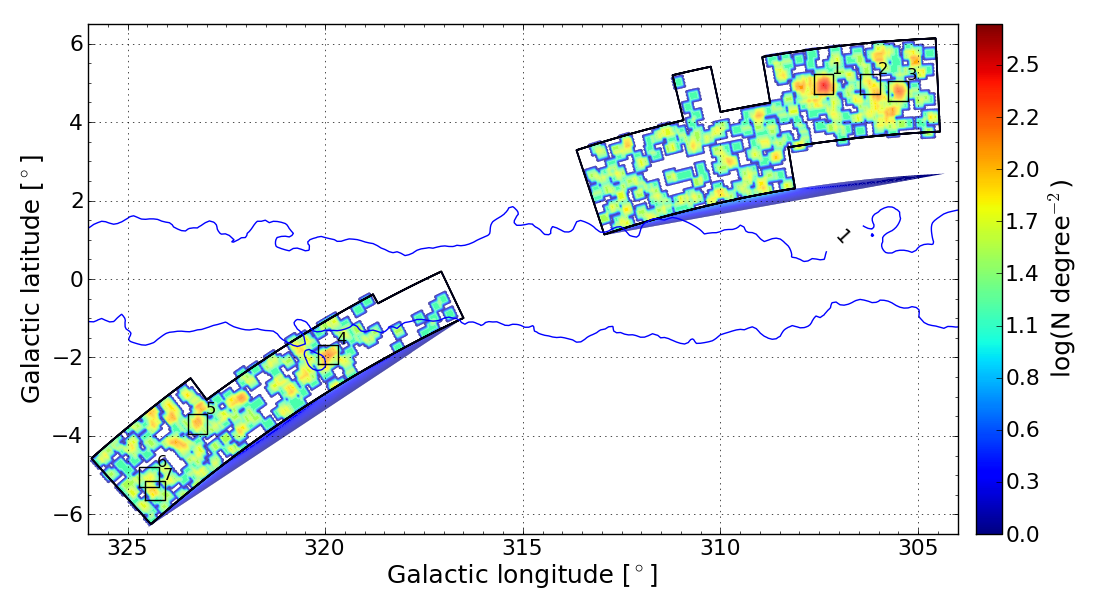}
\includegraphics[width=0.69\textwidth]{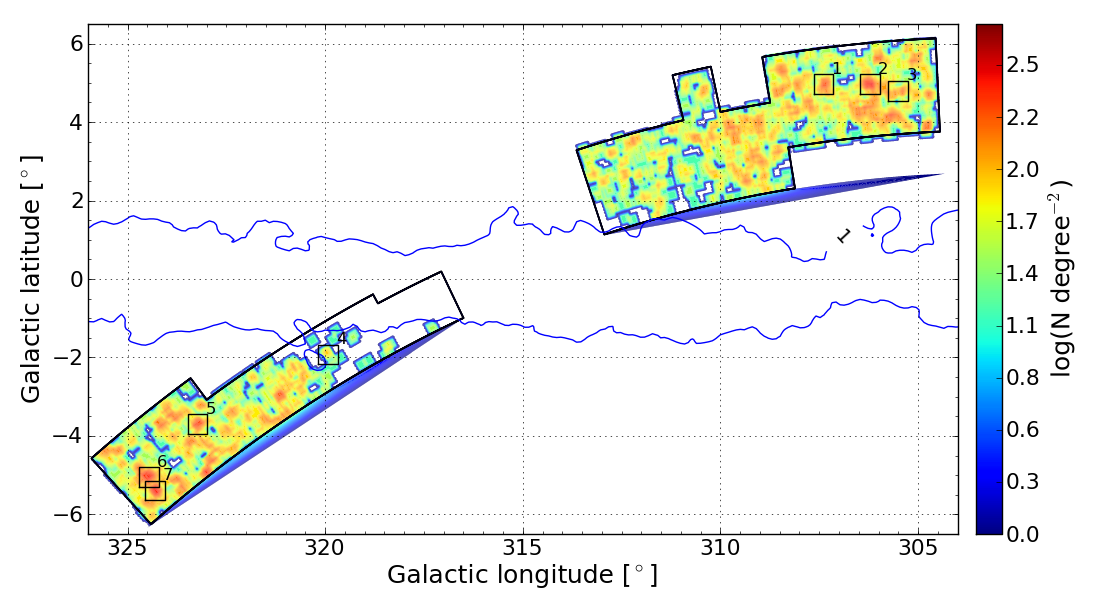}
\caption[Galaxy density counts in magnitude slices]{Galaxy density
  counts in two magnitude slices. Top panel: the distribution for
  galaxies complete to $K_{s}^o<13\fm5$.  Bottom panel: density
  contours for galaxies in the magnitude range $13\fm5\leq
  K_{s}^o\leq14\fm8$; the contour marks $A_K <
  1\fm0)$.}\label{density-contour}
\end{center}
\end{figure}

A first assessment of the large-structure and possible identification
of unknown galaxy clusters in the NWS was made by
inspecting the galaxy density contour maps for the above derived
completeness levels. These are presented in Fig.~\ref{density-contour},
subdivided in two magnitude intervals $K_s^o<13\fm5$ and $13\fm5\leq
K_s^o\leq14\fm8$ for a rough differentiation between more local and
distant density enhancements.

There are seven distinct peaks in the number counts of which six have values
ranging from log\,$N/\Box\dgr\sim 2.1 - 2.3$ compared to the mean of
the survey $1.78\Box\dgr$. The two prominent ones in the top panel
(labeled 1 and 3) coincide with the CIZA J1324.7-5736 (Ebeling et
al. 2002) and Cen-Crux cluster cluster (Woudt \& Kraan-Korteweg 2001)
which at $v=5700$\kms\ and $v=6214$\kms\ are known to form part of the
Norma Wall. 

A comparison of the magnitude histograms in $35\arcmin \times
35\arcmin$ squares centered on the density peaks (not shown here)
supports the notion that the other four density peaks (\#2 above the
Galactic Plane, and \#5, 6, 7 below) must be considerably more distant
compared to the CIZA and Cen-Crux histograms. An independent analysis
using photometric redshifts (see Sect.~4) confirms that these peaks
are about 3 times more distant. They might well be connected to the Ara
and Triangulum Australis clusters (both massive X-ray/CIZA clusters),
possibly forming a filament that connects to the Shapley Concentration.

\begin{figure}[!t]
\begin{center}
\includegraphics[scale=0.52]{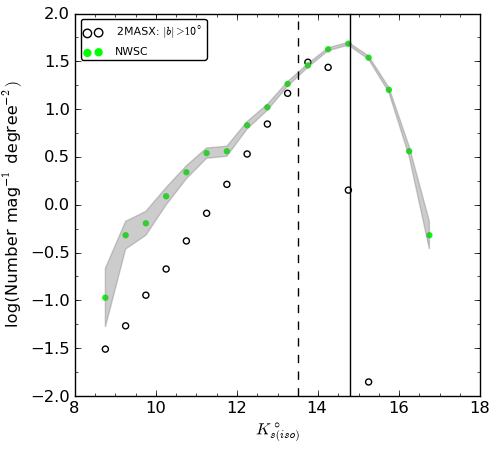}
\caption[Galaxy density counts]{The NWS density counts (lime circles),
  the 2MASX mean density counts (open circles). The completeness
  limits for the NWS ($K_{s(iso)}=14\fm8$) and 2MASX
  ($K_{s(iso)}=13\fm5$) are represented by the solid and dashed
  vertical lines respectively.}\label{NWS-field}
\end{center}
\end{figure}

With no new clusters identified in the Norma Wall, the question arises
what else we can learn about the mass density distribution. To assess
this we compare in Fig.~\ref{NWS-field} the overall number counts to
the mean density counts of the 2MASX all-sky survey outside of the ZOA
($|b| > 10\dgr$) where 2MASX is complete for galaxies with $K_s\le
13\fm5$ (Skrutskie et al. 2006).  The NWS counts reveal a clear
density enhancement over the magnitude range $8\fm5\leq
K_s^o<13\fm5$ and is particularly prominent for $9\fm0\leq
K_s^o<11\fm5$.
This density enhancement is not caused by the CIZA and Cen-Crux
clusters. Excluding the galaxies within a $0\fdg75\times0\fdg75$
regions centered on these clusters affects the overall counts only
minimally, even though the clusters themselves have counts that lie
far above the NWS counts (e.g. Nagayama et al. 2006 for CIZA; Skelton
et al. 2009 Norma). In addition, the bump perseveres whatever sub-area
of the NWS we consider.

We then selected $3^\circ\times3^\circ$ regions from the 2MASX
redshift slices (Jarrett 2004) covering different environments and
redshift intervals to gauge what kind of structures can reproduce the
shape of the NWS counts. They include low-density regions, filamentary
features connecting clusters, and wall-like features feeding into
prominent clusters (but not cluster cores) at distances corresponding
to the surroundings such as Virgo, Pavo, Norma, Vela and even Shapley.

The best correspondence was realized with wall-like structures at the
approximate GA distance -- with a slight improvement for the counts
around $K_s^o \ga 13^{\rm m}$ if the line-of-sight also cuts through a
higher redshift structure (like Shapley). The counts clearly are
incompatible with more local overdensities (e.g. around Virgo) or very
distant filaments (e.g. Shapley surroundings), nor clusters
themselves.

In summary, the survey revealed no previously unknown clusters, but a
clear excess in galaxies at the distance range of the Great Attractor
that extends over the whole NWS area.
 
\section{Assessment of the 3-D distribution}

Localizing the density enhancement in space would be easier if the
distances to the galaxies were known.  However, only few redshifts
exist in the literature for these highly obscured, previously mostly
unknown galaxies. A search in NED found 128 (2.9\%) redshifts --
predominantly for galaxies at the lowest extinction levels. As a first
proxy we therefore started working with photometric redshifts.  This
is hard in the ZOA due to the reddening effect of the selective
absorption. Photometric redshifts were kindly determined for us by
T. Jarrett based on his refined NIR phot-z estimator (Jarrett, in prep.).

A comparison with the 128 published spectroscopic redshifts in common
found the errors of the photometric redshifts to be large
(30-40\%). This resulted in the signature of large-scale structures to
be smeared out (galaxies in the GA region with $0.02 < z_{phot} <
0.03$ are spread over $0.005 < z_{phot} < 0.04$) -- and skewed towards
higher redshifts. This should be taken into account when interpreting
the following results.

\begin{figure}[!htb]
\begin{center}
\includegraphics[scale=0.52]{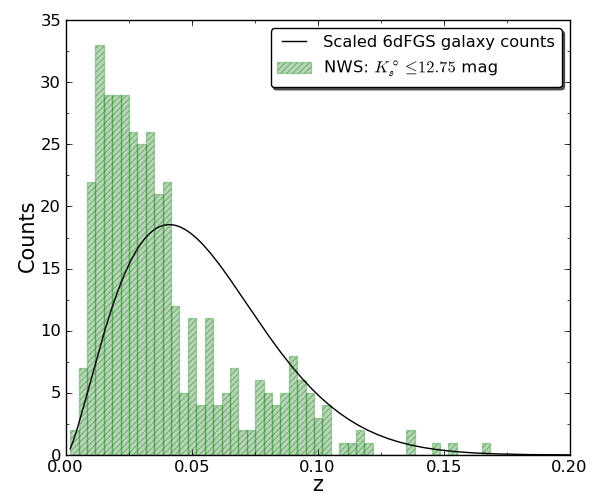}
\caption[Photometric redshift distribution]{Photometric redshift
  histogram for NWS galaxies brighter than $K_{s(iso)}=12\fm75$ with
  the 6dFGS luminosity function scaled to the displayed NWS
  counts.}\label{nws-z-dist}
\end{center}
\end{figure}

Figure~\ref{nws-z-dist} displays the redshift distribution as derived
from phot-z for the NWS galaxy sample delimited by $K^o_s <
12\fm75$. This magnitude limit was imposed to allow a direct
comparison with the 6dF Galaxy Survey (6dFGS; Jones et al. 2006) which
has the same completeness limit. Their luminosity function is also
plotted in Fig.~\ref{nws-z-dist}, but scaled to the NWS counts. It is
given here as a reference for the expected smooth distribution as
derived over a large volume of space.

The redshift histogram shows a prominent broad excess in the phot-z
range $0.005<z_{phot}<0.04$ compared to the 6dFGS field counts. As
argued above, this excess corresponds to the Norma Wall redshift
range, with the broad width originating from the use of photometric
redshift. Hence the histogram in Fig.~\ref{nws-z-dist} provides
independent evidence that the density enhancement observed in
Fig.~\ref{NWS-field} is caused by an excess of galaxies at the GA
distance.

It is worthwhile pointing out that the overall height of the
6dFGS-distribution function would be considerably lower if it were
scaled to counts that exclude the GA overdensity and be more typical
of an average volume. With a lowered 6dFGS curve, however, the second
peak that extends from $0.075 \la z_{phot} \la 0.1$ would stand out
also quite succinctly. In fact, this second peak is suggestive of an
overdensity that corresponds to the high-density peaks identified in
the lower panel of Fig.~\ref{density-contour} -- and the adjacent Ara
and Triangulum Australis clusters -- making the tantalizing link with
the Shapley Concentration even more probable.

\section{Mass estimate of the GA Wall overdensity}

To derive an actual mass estimate of the overdensity we adopt the
extent of the Norma Wall to be $\approx100^\circ$ $(84.5h^{-1}$Mpc)
across the sky (Radburn-Smith et al. 2006), with a mean
cross-sectional radius of $2h^{-1}$ Mpc as suggested to be typical for
filaments (Colberg et al. 2005).  We assume the wall-like feature to
be homogeneously filled with the here observed galaxy density. We then
quantify the density enhancement by fitting a luminosity function to
the observed overdensity. This allows a derivation of the excess luminosity density from integrating over the luminosity function. A suitable
$M\L$ ratio will then yield a mass estimate for the above defined
volume of the Norma Wall.

\begin{figure}[!htb]
\begin{center}
\includegraphics[width=0.31\textwidth,height=0.31\textwidth]{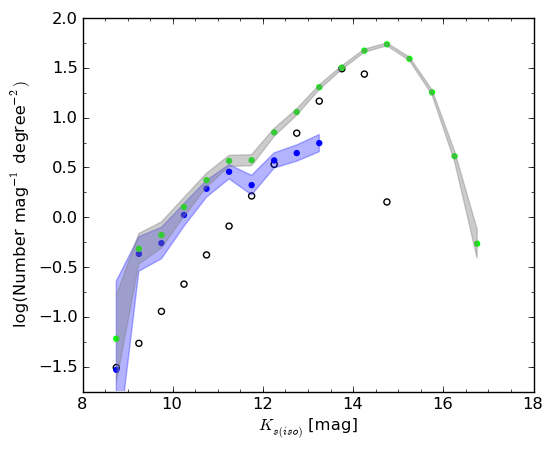}
\includegraphics[width=0.31\textwidth,height=0.31\textwidth]{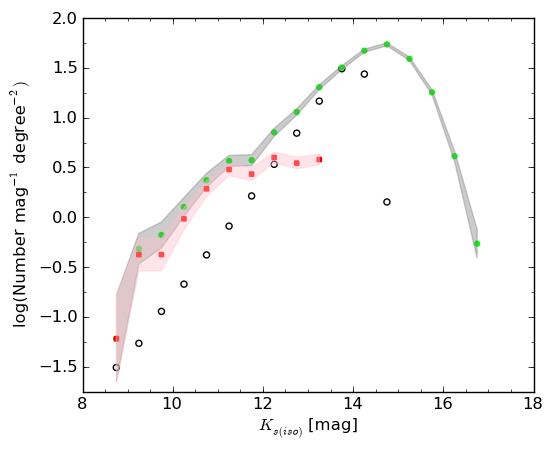}
\includegraphics[width=0.31\textwidth,height=0.31\textwidth]{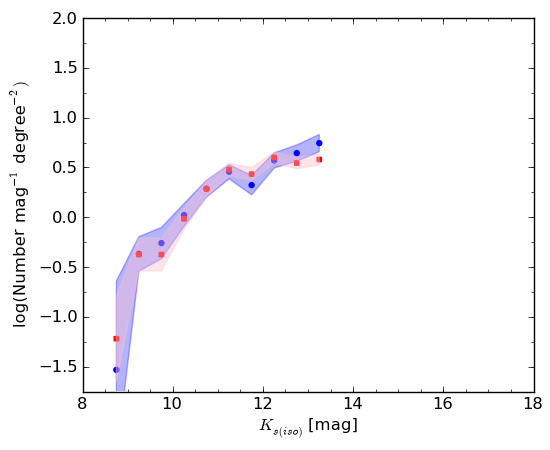}
\caption[Galaxy overdensity counts]{Galaxy overdensity counts. Left panel: 
  The NWS counts (lime circles), the mean 2MASX density counts (open
  circles) and the difference between the two, i.e. the excess density
  counts (blue circles). Middle panel: The all survey area NWS
  counts (lime circles), the mean 2MASX density counts (open circles)
  and the density counts for galaxies in the redshift range
  $0.005<z_{phot}<0.04$. Right panel: The excess density counts (blue
  circles) calculated as the difference between the NWS counts and the
  2MASS mean counts and the density counts for galaxies in the
  redshift range $0.005<z_{phot}<0.04$.}\label{LF1}
\end{center}
\end{figure}

The density enhancement is quantified by two different methods. The
first is based on the total NWS number counts from which the 2MASX
whole sky counts are subtracted up to the common completeness limit
$K^o_s = 13\fm5$. The resulting curve for the excess counts is
displayed in the left panel of Fig.~\ref{LF1}. The second method is
based on galaxies that have phot-z values that localize the galaxies
in the Great Attractor, i.e.  galaxies with
$0.005<z_{phot}<0.04$. This is plotted in the middle panel. The
right-hand panel shows the two resulting curves on the same plot,
emphasizing the superb agreement obtained by the two completely
independent methods.

In a next step we determined the luminosity density of the excess
galaxies by fitting a Schechter (1976) luminosity function to the
above derived curves.  To convert apparent magnitudes to absolute
magnitudes the excess galaxies in the two distributions in
Fig.~\ref{LF1} are assumed to be at the distance of the Norma cluster
i.e. a mean of $z=0.016$ (4844\kms; Woudt et al. 2008). This is a
simplification as we know the Norma Wall to extend to slightly higher
redshifts as it crosses from Norma to the CIZA, Cen-Crux then Vela
clusters.

The parameter sets of the resulting luminosity functions lie well
within the range of what typically is found for $K$-band cluster and
field luminosity functions (Riad 2010). The excess counts have a
steeper faint end, but brighter characteristic magnitude compared to
the phot-z GA galaxies. This is probably due to the degeneracy in the
value of $M^*$ and $\alpha$ of the Schechter function. Despite these
variations the resulting luminosity densities derived from integrating
over the luminosity function are consistent ($\pm10$\%).

\begin{figure}
\begin{center}
\includegraphics[scale=0.48]{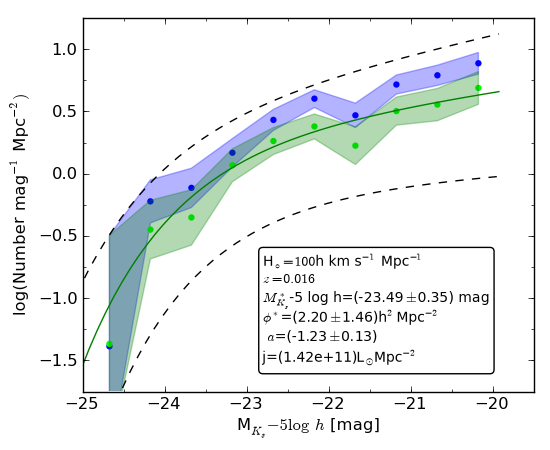}
\caption[Luminosity function for the NWS overdensity excluding the
CIZA J1324.7-5736 and Cen-Crux clusters]{The blue dots are the density
  distribution for the observed overdensity in the NWS defined as the
  excess between the NWS counts density and the 2MASX mean
  density. The green circles represent the excess density counts
  excluding the CIZA J1324.7-5736 and Cen-Crux density counts. The
  Schechter function fitted to the excess distribution excluding the
  clusters is plotted with the green solid line.}\label{LFnocluster}
\end{center}
\end{figure}

To get a better feel for the overall uniform density enhancement we
exclude the clusters in first instance. As shown with
Fig.~\ref{LFnocluster}, which shows both, the resulting LF is slightly
shallower and the luminosity density with $j_K =
1.42\times10^{11}$L$_\odot$ Mpc$^{-2}$ a bit lower (64\%). To convert this
into a mass estimate, we adopt the mass-to-light ratio from Rines et
al. (2004) for regions within 1-10 $h^{-1}$ Mpc of a galaxy
cluster. The Norma wall is a rich structure with a number of known
clusters embedded in it. This makes the use of their value of
$M/L=53\,h$ M$_\odot/$L$_\odot$ a reasonable assumption. 

Filling a cylindrical volume of 84.5$h^{-1}$Mpc length and a radius of
2$h^{-1}$Mpc yields an excess mass in the Norma Wall of
$\approx2.5\times10^{15}h^{-1}$ M$_\odot$ -- excluding clusters.
The clusters Norma, Pavo II, CIZA J1324.7-5736, Cen-Crux and Abell
S0639 contribute an additional mass of $\approx1.6\times10^{15}h^{-1}$
M$_\odot$ (Riad 2010, Riad et al. 2011). This gives a combined mass of
$\approx4.1\times10^{15}h^{-1}$ M$_\odot$ for the Norma Wall as confined by
the considered volume.

\section{Conclusions}

The homogeneous galaxy distribution over the survey footprint is
suggestive of a continuous structure across the GP, i.e. the so-called
Norma Wall. The Norma Wall survey did not uncover any previously
unknown clusters that form part of the GA overdensity. Assuming the
Great Attractor, respectively the Norma Wall, to be a cylindrical
filament of 84.5$h^{-1}$ Mpc in length with a radius of 2$h^{-1}$
Mpc that is filled uniformly by the mean overdensity derived at the
approximate GA distance, and including the Norma, Pavo II, CIZA
J1324.7-5736, Cen-Crux and Abell S0639 clusters, the total mass excess
in galaxies amounts to $\sim 4\times 10^{15}h^{-1}\msun$.  This mass
is in good agreement with recent estimates for the mass of the Norma
Wall from optical, X-ray and HI observations (Radburn-Smith et
al. 2006; Kocevski et al. 2007; Stavely-Smith et al 2000) but lower
than the original estimate by Lynden-Bell et al. (1988).\\~\\

\noindent {\bf Acknowledgements. ---}
\begin{small}
{RCKK and IFR acknowledge financial support from the National Research Foundation through a mobility grant from the South African -
Japanese (NRF/JSPS) bilateral grant in Astronomy. IFR was supported throughout his PhD studies by the University of Khartoum, South African National
and the Stichting Steunfonds Soedanese Studenten. We are grateful to T. Jarrett for his derivation of the photometric redshifts.}
\end{small}


\begin{thebibliography}{}
\begin{small} 
\bibitem[{XX}(xxxx)]{}
Colberg J.M., Krughoff K.S., \& Connolly A.J. 2005, MNRAS 359, 272
\bibitem[{XX}(xxxx)]{}
Ebeling H., Mullis C.R. \& Tully R.B. 2002, ApJ 580, 774
\bibitem[{XX}(xxxx)]{}
Jarrett T.H. 2004, PASP, 21, 396
\bibitem[{XX}(xxxx)]{}
Jarrett T.H., Chester T., Cutri R. et al. 2000, AJ 119, 2498 [2MASX]
\bibitem[{XX}(xxxx)]{} 
Jones D.H, Peterson B.A, Colless M., \& Saunders W. 2006, MNRAS 369, 25
\bibitem[{XX}(xxxx)]{} 
Kocevski D.D., Ebeling H., Mullis C.R. \& Tully R.B. 2007, ApJ 662, 224
\bibitem[{XX}(xxxx)]{} 
Kraan-Korteweg, R.C. 2005, RvMA, 18, 48
\bibitem[{XX}(xxxx)]{}
Kraan-Korteweg R.C. \& Lahav O. 2000, A\&ARv 10, 211
\bibitem[{XX}(xxxx)]{}
Lynden-Bell D., Faber S.M., Burstein D. et al. 1988, ApJ 326, 19
\bibitem[{XX}(xxxx)]{} 
Nagayama T., Woudt P.A, Wakamatsu K., et al. 2006, MNRAS 368, 534
\bibitem[{XX}(xxxx)]{} 
Radburn-Smith D.J., Lucey J.R., Woudt P.A., Kraan-Korteweg R.C. \& Watson F.G.
2006, MNRAS 369, 1131
\bibitem[{XX}(xxxx)]{} 
Riad I.F. 2010, PhD thesis, University of Cape Town
\bibitem[{XX}(xxxx)]{} 
Riad I.F., Kraan-Korteweg R.C. \& Woudt P.A. 2010, MNRAS 401, 924
\bibitem[{XX}(xxxx)]{} 
Rines K., Geller M.J., Diaferio A., Kurtz M.J. \& Jarrett T.H. 2004, AJ 128, 1078
\bibitem[{XX}(xxxx)]{} 
Schechter P. 1976, ApJ 203, 297
\bibitem[{XX}(xxxx)]{} 
Skelton R.E., Woudt P.A. \& Kraan-Korteweg R.C. 2009, MNRAS 396, 2367
\bibitem[{XX}(xxxx)]{} 
Skrutskie M.F., Cutri R.M. Stiening, R. et al. 2006, AJ, 131, 1163
\bibitem[{XX}(xxxx)]{} 
Staveley-Smith L., Juraszek S., Henning P.A., Koribalski B.S. \& Kraan-Korteweg R.C. 2000, ASP Conf. Ser. 218, 207
\bibitem[{XX}(xxxx)]{} 
Wakamatsu K. 2011, these proceedings
\bibitem[{XX}(xxxx)]{} 
Williams W.L., Woudt P.A. \& Kraan-Korteweg R.C. 2011, these proceedings
\bibitem[{XX}(xxxx)]{} 
Woudt P.A. \& Kraan-Korteweg R.C. 2001, A\&A 380, 411
\bibitem[{XX}(xxxx)]{} 
Woudt P.A., Kraan-Korteweg R.C., Lucey J., Fairall A.P. \& Moore S.A.W. 2008,
MNRAS 383, 445
\end{small}
\end{thebibliography}
\end{document}